\documentclass[aps,prd,twocolumn,showpacs,amsmath,amssymb,superscriptaddress]{revtex4}
\usepackage{graphicx}
\usepackage{dcolumn}
\usepackage{bm}

\begin{document}
\def\slashchar#1{\setbox0=\hbox{$#1$}           
   \dimen0=\wd0                                 
   \setbox1=\hbox{/} \dimen1=\wd1               
   \ifdim\dimen0>\dimen1                        
      \rlap{\hbox to \dimen0{\hfil/\hfil}}      
      #1                                        
   \else                                        
      \rlap{\hbox to \dimen1{\hfil$#1$\hfil}}   
      /                                         
   \fi}                                         %
\def\etmiss{\slashchar{E}_T}			%
\def\htmiss{\slashchar{H}_T}			%
\def\ptmiss{\slashchar{p}_T}			%
\newcommand\lsb{\tilde{b}_1}
\newcommand\alsb{\bar{\tilde{b}}}
\newcommand\lntr{\tilde{\chi}_{1}^0}
\hyphenation{ALPGEN}


%

%
\title{Search for scalar bottom quarks and
third-generation leptoquarks 
in \boldmath{$p\overline{p}$} collisions at \boldmath{$\sqrt{s}$} = 1.96 TeV}

%
%
\affiliation{Universidad de Buenos Aires, Buenos Aires, Argentina}
\affiliation{LAFEX, Centro Brasileiro de Pesquisas F{\'\i}sicas, Rio de Janeiro, Brazil}
\affiliation{Universidade do Estado do Rio de Janeiro, Rio de Janeiro, Brazil}
\affiliation{Universidade Federal do ABC, Santo Andr\'e, Brazil}
\affiliation{Instituto de F\'{\i}sica Te\'orica, Universidade Estadual Paulista, S\~ao Paulo, Brazil}
\affiliation{Simon Fraser University, Vancouver, British Columbia, and York University, Toronto, Ontario, Canada}
\affiliation{University of Science and Technology of China, Hefei, People's Republic of China}
\affiliation{Universidad de los Andes, Bogot\'{a}, Colombia}
\affiliation{Charles University, Faculty of Mathematics and Physics, Center for Particle Physics, Prague, Czech Republic}
\affiliation{Czech Technical University in Prague, Prague, Czech Republic}
\affiliation{Center for Particle Physics, Institute of Physics, Academy of Sciences of the Czech Republic, Prague, Czech Republic}
\affiliation{Universidad San Francisco de Quito, Quito, Ecuador}
\affiliation{LPC, Universit\'e Blaise Pascal, CNRS/IN2P3, Clermont, France}
\affiliation{LPSC, Universit\'e Joseph Fourier Grenoble 1, CNRS/IN2P3, Institut National Polytechnique de Grenoble, Grenoble, France}
\affiliation{CPPM, Aix-Marseille Universit\'e, CNRS/IN2P3, Marseille, France}
\affiliation{LAL, Universit\'e Paris-Sud, CNRS/IN2P3, Orsay, France}
\affiliation{LPNHE, Universit\'es Paris VI and VII, CNRS/IN2P3, Paris, France}
\affiliation{CEA, Irfu, SPP, Saclay, France}
\affiliation{IPHC, Universit\'e de Strasbourg, CNRS/IN2P3, Strasbourg, France}
\affiliation{IPNL, Universit\'e Lyon 1, CNRS/IN2P3, Villeurbanne, France and Universit\'e de Lyon, Lyon, France}
\affiliation{III. Physikalisches Institut A, RWTH Aachen University, Aachen, Germany}
\affiliation{Physikalisches Institut, Universit{\"a}t Freiburg, Freiburg, Germany}
\affiliation{II. Physikalisches Institut, Georg-August-Universit{\"a}t G\"ottingen, G\"ottingen, Germany}
\affiliation{Institut f{\"u}r Physik, Universit{\"a}t Mainz, Mainz, Germany}
\affiliation{Ludwig-Maximilians-Universit{\"a}t M{\"u}nchen, M{\"u}nchen, Germany}
\affiliation{Fachbereich Physik, Bergische  Universit{\"a}t Wuppertal, Wuppertal, Germany}
\affiliation{Panjab University, Chandigarh, India}
\affiliation{Delhi University, Delhi, India}
\affiliation{Tata Institute of Fundamental Research, Mumbai, India}
\affiliation{University College Dublin, Dublin, Ireland}
\affiliation{Korea Detector Laboratory, Korea University, Seoul, Korea}
\affiliation{SungKyunKwan University, Suwon, Korea}
\affiliation{CINVESTAV, Mexico City, Mexico}
\affiliation{FOM-Institute NIKHEF and University of Amsterdam/NIKHEF, Amsterdam, The Netherlands}
\affiliation{Radboud University Nijmegen/NIKHEF, Nijmegen, The Netherlands}
\affiliation{Joint Institute for Nuclear Research, Dubna, Russia}
\affiliation{Institute for Theoretical and Experimental Physics, Moscow, Russia}
\affiliation{Moscow State University, Moscow, Russia}
\affiliation{Institute for High Energy Physics, Protvino, Russia}
\affiliation{Petersburg Nuclear Physics Institute, St. Petersburg, Russia}
\affiliation{Stockholm University, Stockholm and Uppsala University, Uppsala, Sweden }
\affiliation{Lancaster University, Lancaster LA1 4YB, United Kingdom}
\affiliation{Imperial College London, London SW7 2AZ, United Kingdom}
\affiliation{The University of Manchester, Manchester M13 9PL, United Kingdom}
\affiliation{University of Arizona, Tucson, Arizona 85721, USA}
\affiliation{University of California Riverside, Riverside, California 92521, USA}
\affiliation{Florida State University, Tallahassee, Florida 32306, USA}
\affiliation{Fermi National Accelerator Laboratory, Batavia, Illinois 60510, USA}
\affiliation{University of Illinois at Chicago, Chicago, Illinois 60607, USA}
\affiliation{Northern Illinois University, DeKalb, Illinois 60115, USA}
\affiliation{Northwestern University, Evanston, Illinois 60208, USA}
\affiliation{Indiana University, Bloomington, Indiana 47405, USA}
\affiliation{Purdue University Calumet, Hammond, Indiana 46323, USA}
\affiliation{University of Notre Dame, Notre Dame, Indiana 46556, USA}
\affiliation{Iowa State University, Ames, Iowa 50011, USA}
\affiliation{University of Kansas, Lawrence, Kansas 66045, USA}
\affiliation{Kansas State University, Manhattan, Kansas 66506, USA}
\affiliation{Louisiana Tech University, Ruston, Louisiana 71272, USA}
\affiliation{University of Maryland, College Park, Maryland 20742, USA}
\affiliation{Boston University, Boston, Massachusetts 02215, USA}
\affiliation{Northeastern University, Boston, Massachusetts 02115, USA}
\affiliation{University of Michigan, Ann Arbor, Michigan 48109, USA}
\affiliation{Michigan State University, East Lansing, Michigan 48824, USA}
\affiliation{University of Mississippi, University, Mississippi 38677, USA}
\affiliation{University of Nebraska, Lincoln, Nebraska 68588, USA}
\affiliation{Rutgers University, Piscataway, New Jersey 08855, USA}
\affiliation{Princeton University, Princeton, New Jersey 08544, USA}
\affiliation{State University of New York, Buffalo, New York 14260, USA}
\affiliation{Columbia University, New York, New York 10027, USA}
\affiliation{University of Rochester, Rochester, New York 14627, USA}
\affiliation{State University of New York, Stony Brook, New York 11794, USA}
\affiliation{Brookhaven National Laboratory, Upton, New York 11973, USA}
\affiliation{Langston University, Langston, Oklahoma 73050, USA}
\affiliation{University of Oklahoma, Norman, Oklahoma 73019, USA}
\affiliation{Oklahoma State University, Stillwater, Oklahoma 74078, USA}
\affiliation{Brown University, Providence, Rhode Island 02912, USA}
\affiliation{University of Texas, Arlington, Texas 76019, USA}
\affiliation{Southern Methodist University, Dallas, Texas 75275, USA}
\affiliation{Rice University, Houston, Texas 77005, USA}
\affiliation{University of Virginia, Charlottesville, Virginia 22901, USA}
\affiliation{University of Washington, Seattle, Washington 98195, USA}
\author{V.M.~Abazov} \affiliation{Joint Institute for Nuclear Research, Dubna, Russia}
\author{B.~Abbott} \affiliation{University of Oklahoma, Norman, Oklahoma 73019, USA}
\author{M.~Abolins} \affiliation{Michigan State University, East Lansing, Michigan 48824, USA}
\author{B.S.~Acharya} \affiliation{Tata Institute of Fundamental Research, Mumbai, India}
\author{M.~Adams} \affiliation{University of Illinois at Chicago, Chicago, Illinois 60607, USA}
\author{T.~Adams} \affiliation{Florida State University, Tallahassee, Florida 32306, USA}
\author{E.~Aguilo} \affiliation{Simon Fraser University, Vancouver, British Columbia, and York University, Toronto, Ontario, Canada}
\author{G.D.~Alexeev} \affiliation{Joint Institute for Nuclear Research, Dubna, Russia}
\author{G.~Alkhazov} \affiliation{Petersburg Nuclear Physics Institute, St. Petersburg, Russia}
\author{A.~Alton$^{a}$} \affiliation{University of Michigan, Ann Arbor, Michigan 48109, USA}
% ...
%
\collaboration{The D0 Collaboration\footnote{with visitors from
$^{a}$Augustana College, Sioux Falls, SD, USA,
$^{b}$The University of Liverpool, Liverpool, UK,
$^{c}$SLAC, Menlo Park, CA, USA,
$^{d}$ICREA/IFAE, Barcelona, Spain,
$^{e}$Centro de Investigacion en Computacion - IPN, Mexico City, Mexico,
$^{f}$ECFM, Universidad Autonoma de Sinaloa, Culiac\'an, Mexico,
and
$^{g}$Universit{\"a}t Bern, Bern, Switzerland.%
}} \noaffiliation
\vskip 0.25cm

%
%
%

\date{May 12, 2010}

\begin{abstract}
We report the results of a search for pair production of 
 scalar bottom quarks ($\lsb$) and scalar 
third-generation leptoquarks ($LQ_3$) in 5.2 fb$^{-1}$ of $p\bar{p}$ 
collisions at
the D0 experiment of the Fermilab Tevatron 
Collider. 
Scalar bottom quarks are 
assumed to decay to a neutralino ($\lntr$) and a $b$ quark, and 
we set 95\% C.L. lower 
limits on their production in the ($m_{\lsb}, m_{\lntr}$) mass 
plane such as
 $m_{\lsb}>247$~GeV for $m_{\lntr}=0$ and $m_{\lntr}>110$~GeV for 
$160<m_{\lsb}< 200$~GeV.
The leptoquarks are assumed to decay to a tau neutrino and a 
$b$ quark, and we set a 95\% C.L. lower limit 
of
247~GeV on the mass of a charge-1/3 third-generation scalar leptoquark.  
%

\end{abstract}


\maketitle

\newpage

%
The standard model (SM) offers an accurate description of current 
experimental data in high energy physics but it is believed
to be embedded in a more general theory.
In particular, extensions of the SM to higher mass 
scales have been proposed that predict the existence of new 
particles and phenomena which can be searched for at
the Tevatron.
 
Supersymmetric (SUSY) models provide an extension of the SM that 
resolves the ``hierarchy problem" by introducing supersymmetric 
partners to 
the known fermions and bosons \cite{susy}. The 
supersymmetric quarks (squarks) are mixtures of the
states $\tilde{q}_L$ and $\tilde{q}_R$, the superpartners of 
the 
SM quark helicity states. The theory permits a mass difference
between the squark mass eigenstates, $\tilde{q}_1$ and  $\tilde{q}_2$, 
and allows the possibility that the lighter states of top and bottom 
squarks have masses smaller than the squarks of the first 
two generations. 
In this analysis we consider the region of SUSY parameter space where
the only decay of the lighter bottom squark is
$\tilde{b}_1\rightarrow b\lntr$, with
$m_b+m_{\lntr} <
m_{\tilde{b}_1}<m_t+m_{\tilde{\chi}_1^-}$,
and the neutralino $\lntr$ and chargino $\tilde{\chi}_1^\pm$ are the 
lightest SUSY partners of the electroweak and Higgs bosons. 
This analysis is interpreted within the framework of 
 the minimal 
supersymmetric standard model (MSSM) with $R$-parity~\cite{mssm}  
conservation, and under the hypothesis that the lightest, and consequently
stable, SUSY  particle is the $\lntr$. 
We therefore search for  
$p\bar{p} \rightarrow \tilde{b}_1\bar{\tilde{b}}_1 \rightarrow b\lntr\bar{b}\lntr$.

Leptoquarks are hypothesized fundamental particles that  
 have color, electric charge,
and both lepton and baryon quantum numbers. They appear in many 
extensions 
of the SM including extended gauge theories, composite models, and 
SUSY with $R$-parity violation~\cite{lqtheory}.
Current models suggest that leptoquarks of each of
the three generations 
should decay to the corresponding generation of SM
leptons and quarks to avoid introducing unwanted flavor changing neutral
currents.
Charge-1/3 third-generation leptoquarks would decay to 
$b\nu$ with branching fraction $B$ or to 
$t\tau$ with branching fraction $1-B$.

We report on a search for the production of pairs 
of 
bottom squarks and third-generation scalar leptoquarks in data 
collected by the D0 Collaboration at the Fermilab Tevatron 
Collider. For both searches, the signature is 
defined to be two $b$-jets and 
missing 
transverse energy ($\etmiss$) from the escaping neutrinos or 
neutralinos. 
This topology is identical to that for
$p\bar{p}\rightarrow ZH \rightarrow \nu\bar{\nu}+b\bar{b}$ 
production, and the two 
analyses are based on the same data and 
selection criteria~\cite{zhpaper}. 
Bottom squark or leptoquark pairs are expected to be produced mainly 
through
$q\bar{q}$ annihilation or $gg$ fusion, with identical leading order 
QCD production cross sections.
We use the next-to-leading order (NLO) cross 
sections calculated by {\sc prospino} 2.1 
for both 
bottom squark~\cite{prospino} and 
leptoquark~\cite{Kramer:1997hh}  
pair production, and found them to agree to better than 3\%.
Previous measurements excluded bottom squark masses $m_{\lsb} <
222$~GeV for a massless neutralino~\cite{sboldD0}, 
as well as charge-1/3 third-generation scalar
leptoquark masses $m_{LQ}<229$~GeV for $B=1$~\cite{lqold}.

The D0 detector \cite{d0nim} consists of layered 
systems surrounding the interaction point. The momenta of charged
particles and the location of the interaction vertices are 
determined using a silicon microstrip tracker and a central 
fiber tracker immersed in the magnetic field of a 2 T solenoid. Jets, 
electrons, and tau leptons
are reconstructed using the tracking information and the pattern of energy 
deposits
in three uranium/liquid-argon calorimeters located outside
the tracking system with a central calorimeter covering pseudorapidity
$|\eta| < 1.1$, and two end calorimeters housed in separate
cryostats covering the regions up to $|\eta| \approx 4.2$.
Jet reconstruction uses a cone algorithm \cite{blazey} with radius
${\cal R} = \sqrt{(\Delta y)^2+(\Delta \phi)^2} = 0.5$ in 
rapidity ($y$) and 
azimuth ($\phi$).
Muons are identified through the association of tracks
with hits in the muon system, which is outside of the calorimeter 
and consists of drift tubes
and scintillation counters before and after 1.8 T iron toroids. 
The $\etmiss$ is determined from the negative of the vector sum of
the transverse components of the energy deposited in the
calorimeter and the transverse momenta $p_T$ of detected muons.
The jet energies are calibrated using transverse energy
balance in events with photons and jets and this calibration is 
propagated to the value of $\etmiss$. 

The data were recorded 
using triggers based on jets and the $\etmiss$ in the event. In 
addition 
to requirements on $\etmiss$ and 
jet energy, the vector sum of the transverse energies of 
all jets, 
defined as  $\htmiss \equiv
|\sum_{\text{jets}}\vec{p}_T|$, the scalar sum of the $p_T$ of the jets 
($H_T$), and the angle $\alpha$
between the two leading jets in the transverse plane, are 
also used for triggering.
Typical requirements are  $\etmiss > 25$~GeV, $\htmiss > 25$~GeV, 
$H_T >$ 50~GeV, and $\alpha < 169^\circ$. 
After imposing quality requirements,
the data correspond to an integrated luminosity of 
5.2~fb$^{-1}$. The previous D0 publications~\cite{sboldD0,lqold} 
used a subset of this data sample, and are superseded by the results
obtained in this Letter.
 
Monte Carlo (MC) samples for $200< m_{LQ}<280$~GeV, and for  
($\lsb,\lntr$) pairs with 
$80<m_{\lsb}<260$~GeV and $m_{\lntr}<120$~GeV, are 
generated with {\sc pythia}~\cite{Pythia}. 
Backgrounds from SM processes with significant $\etmiss$ 
are estimated using MC. 
The most important backgrounds are from $W/Z$ bosons 
produced in association with 
jets, with leptonic decays such 
as $Z \rightarrow \nu\bar{\nu}$ and $W\rightarrow e\nu$,
and processes with $t\bar{t}$ and single top quark production. The cross 
sections used 
to estimate these contributions to the background are obtained from 
 \cite{mc_xsec} and \cite{mc_xsec1}. 
At the parton level, vector boson pair production and the single-top 
quark events 
are generated with {\sc pythia} and {\sc comphep}~\cite{CompHEP}, 
respectively, while  {\sc alpgen}~\cite{Alpgen} is used 
for all other samples.
All MC events are then processed with {\sc pythia}, 
which performs parton showering and hadronization. 
The resulting samples are 
processed using a  {\sc geant}~\cite{GEANT} simulation of the D0 
detector. 
To model the effects of multiple interactions and
detector noise, data from random $p\bar{p}$ crossings are 
overlaid on MC events.
The {\sc cteq6l1} parameterization~\cite{PDF:CTEQ6L1} is 
used for all parton density functions (PDF). Instrumental 
background 
comes mostly from multijet processes with $\etmiss$ arising from 
energy mismeasurement.
This background, which we label MJ, dominates the low $\etmiss$ region 
and is modeled using data.

A signal sample and a sample used 
to model the MJ background are selected.
We select events with two or three jets with 
$\vert\eta\vert<2.5$ and  $p_T>20$~GeV, and require that
the interaction vertex has at least three tracks and is reconstructed 
within $\pm$40~cm 
of the center of the detector along the beam direction so that the tracks
are within the geometric acceptance of the silicon tracker.
As the leading highest $p_T$ jets in the signal events are assumed  
to originate from decays of $b$ quarks, we require that at least 
two jets, including the leading jet, 
have at least two tracks pointing to the primary vertex
in order to apply b-tagging algorithms.
We also require the two leading jets satisfy $\alpha < 165^\circ$. 
To reduce the contribution from $W\rightarrow l\nu$ decays, 
we veto events with isolated electrons or  
muons with $p_T>15$~GeV, as well as tau leptons that decay hadronically
to a single charged particle with 
$p_{T}>12$~GeV when there is no associated electromagnetic 
cluster or $p_{T}>10$~GeV if there is such a cluster~\cite{tauid}. 
To suppress the MJ background, we require $\etmiss>40$~GeV and
$\etmiss$ significance ${\cal S} > 5$ \cite{metsig}.
We also remove events when the direction of the $\etmiss$ overlaps 
with a jet in 
$\phi$ by requiring  
$\etmiss/$GeV $> 80 - 40 \times {\Delta\phi}_{\text{min}}(\etmiss,$~jets),
where
$\Delta{\phi}_{\text{min}}(\etmiss$,~jets) denotes the minimum of the 
angles 
between the $\etmiss$ and any of the selected jets.

\begin{figure}[ht]
\begin{center}
\includegraphics[scale=0.40]{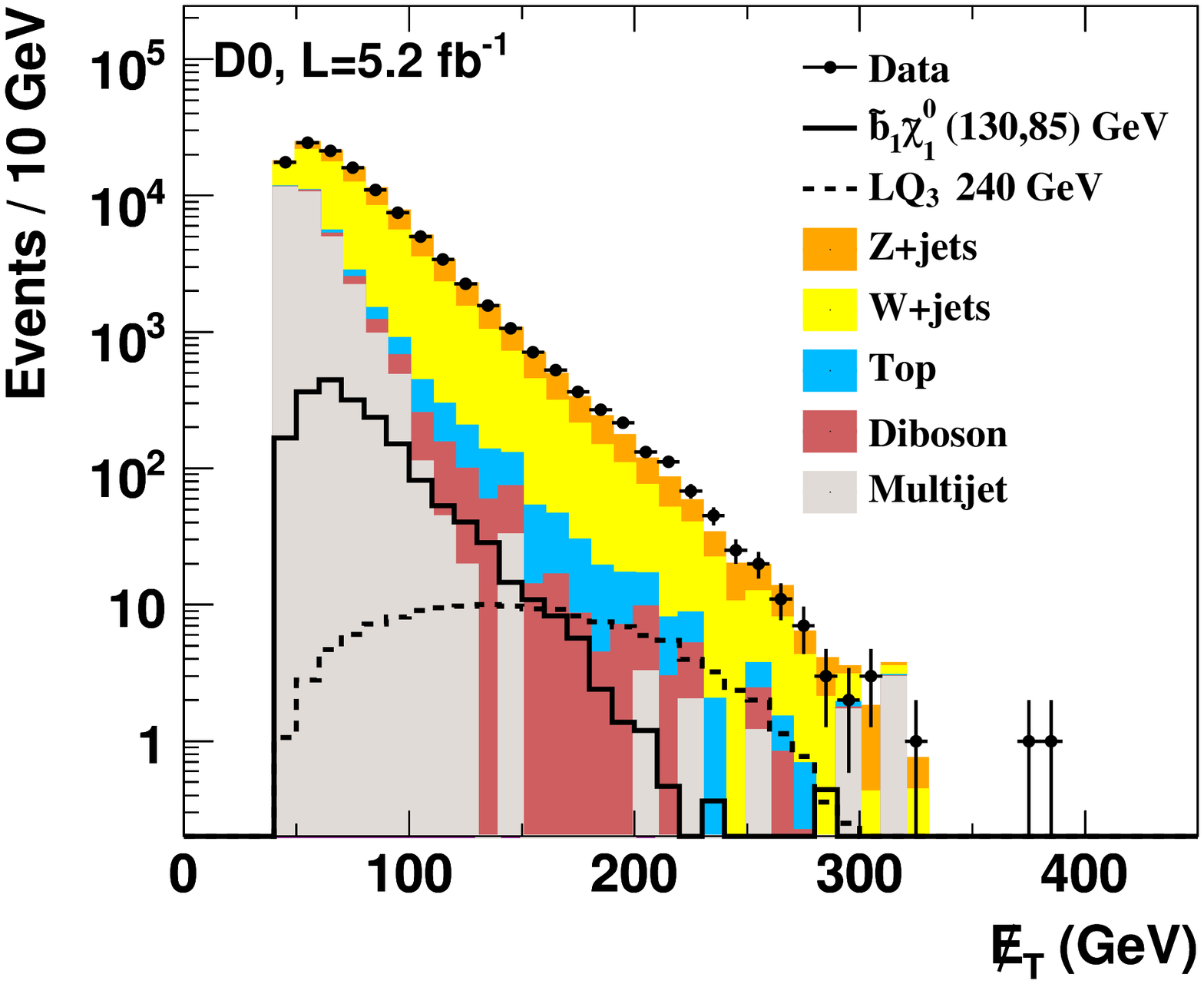}
\end{center}
\caption{\label{fig:met_and_mht_pretag} (color online). 
The $\etmiss$ distribution 
before $b$-tagging. The points with the error bars represent data while 
the shaded histograms show the contributions from 
background processes. Signal distributions with 
($m_{\lsb}$,$m_{\lntr}$)=(130,85)~GeV
and $m_{LQ}=240$~GeV are
 shown as solid and dashed lines, respectively. 
}
\end{figure}

The contribution from multijet processes is determined using the 
techniques described in \cite{zhpaper}.
For signal events, the  
direction of $\etmiss$ tends to be aligned with the missing
track transverse momentum, $\ptmiss$, defined as the negative of the 
vectorial sum of 
the $p_T$ of the charged particles. A strong correlation of this kind
is not expected in multijet events, where $\etmiss$ originates mainly from 
mismeasurement of jet energies in the calorimeter. We exploit this 
difference by  
requiring ${\cal D} < \pi/2 $ for signal,
where ${\cal{D}}$ is the azimuthal distance between $\etmiss$ and 
$\ptmiss$, $\Delta\phi(\etmiss,\ptmiss)$, and 
use events with ${\cal D} > \pi/2 $
to model the kinematic distributions of the MJ background in the signal 
sample
after subtracting the contribution from SM processes.
The MJ background is normalized before $b$-tagging by requiring 
the number of observed events in data to equal the sum of SM and MJ
contributions 
in the ${\cal D} < \pi/2$ region. The 
signal contribution is assumed to be 
zero. Figure~\ref{fig:met_and_mht_pretag} shows the $\etmiss$ distribution 
and the background contributions from SM and MJ sources after these 
selections.

A neural network (NN) $b$-tagging algorithm~\cite{nnbtag} 
is used
to identify heavy-flavor jets, and reduce the 
SM and MJ backgrounds that are dominated by light flavor jets.  
We apply $b$-tagging 
and use the requirements on the NN output that give
one jet to be tagged with an average efficiency of $\approx$70\% and the 
other with 
an average
efficiency of $\approx$50\%,
where the corresponding probabilities of a light-flavored jet to be 
wrongly identified as a $b$-jet are $\approx $6.5\% and $\approx$0.5\%, 
respectively. 
These conditions are designed to optimize the 
discovery reach for a $\lsb$ and $LQ_3$.

\begin{figure}[th]
\begin{center}
\includegraphics[scale=0.40]{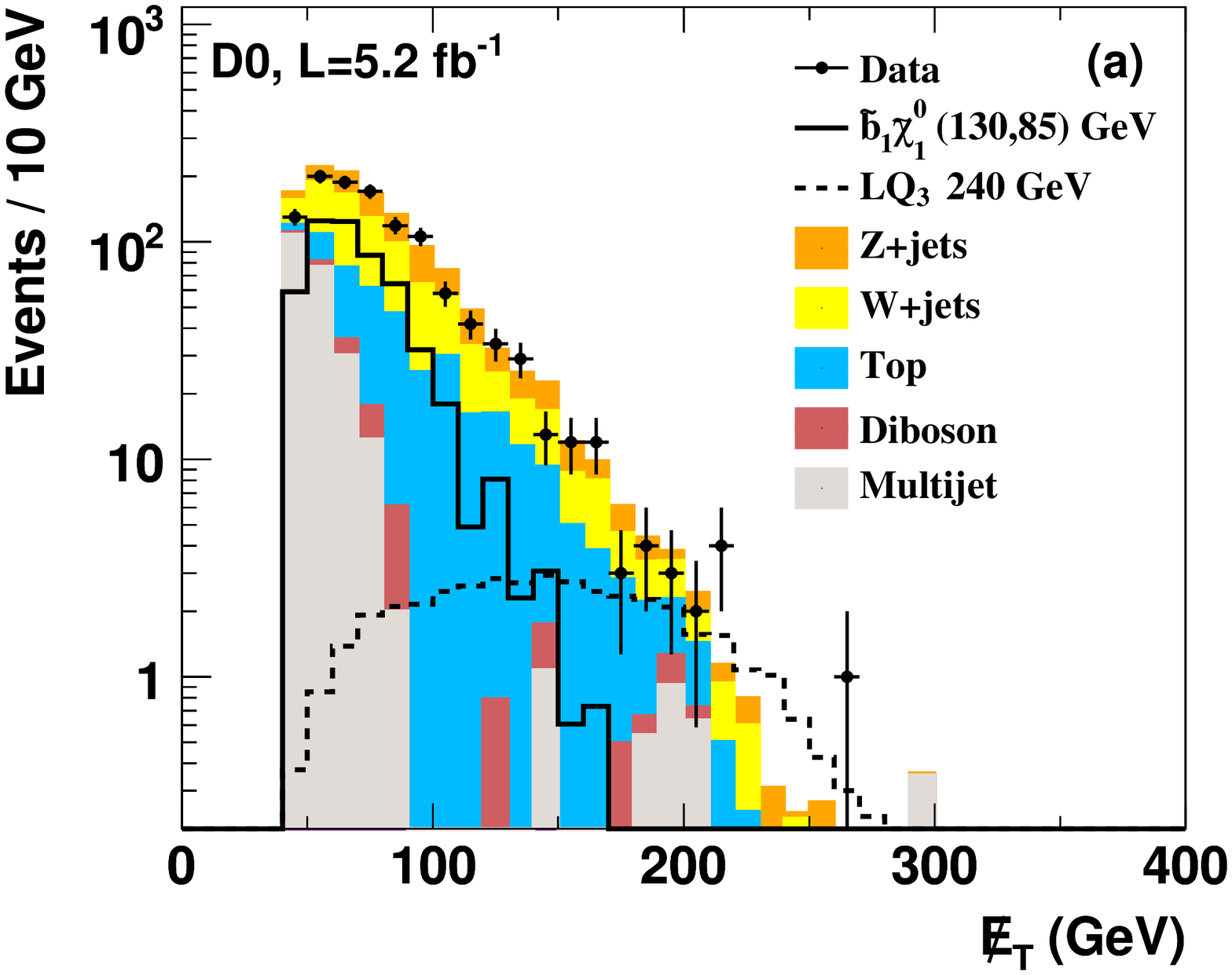} 
\\
\includegraphics[scale=0.40]{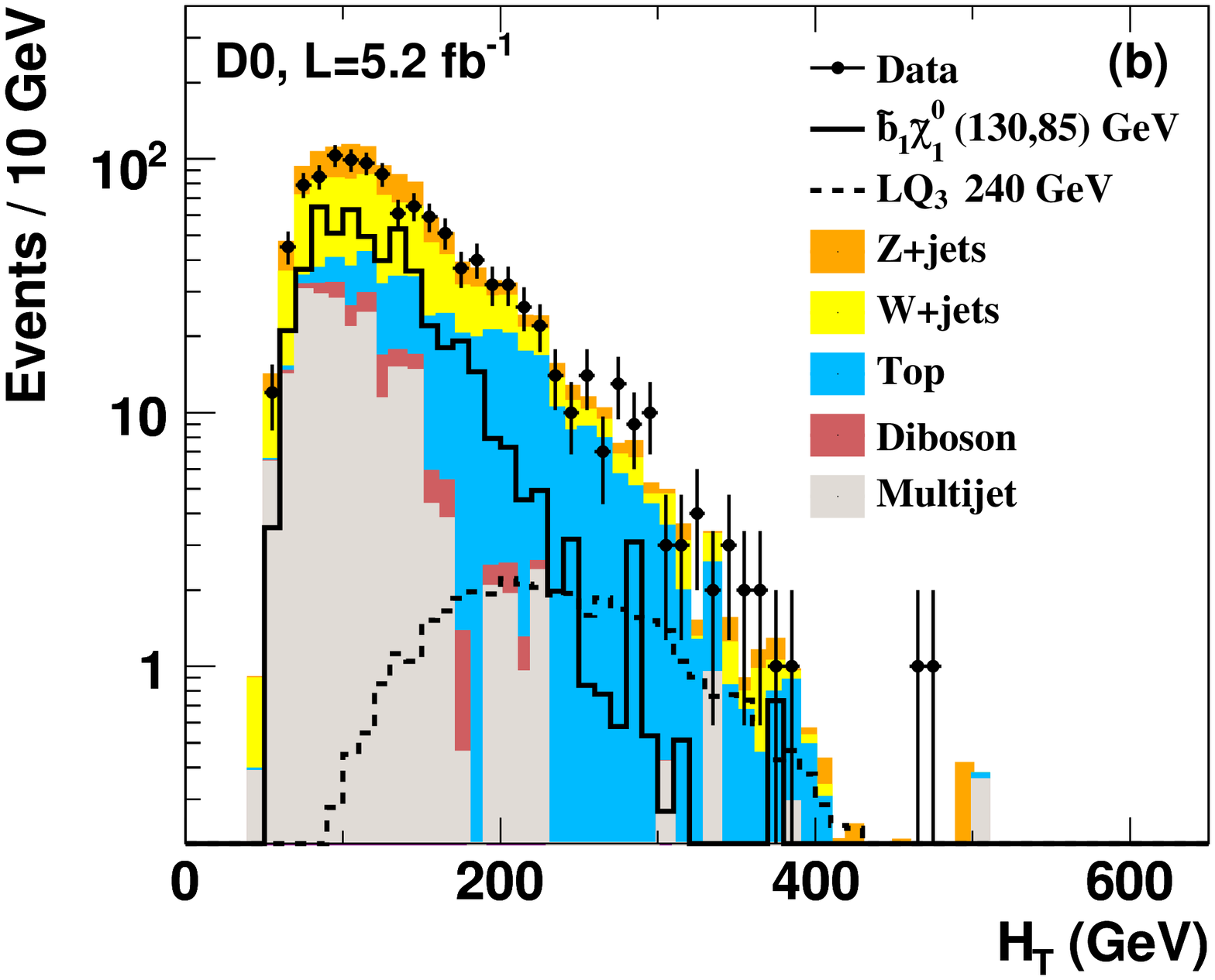}
\end{center}
\caption{\label{fig:L3VTbtag} (color online). The (a) $\etmiss$ 
and (b) $H_T$ distributions after $b$-tagging and additional selections.
The points with the error bars represent data while the shaded histograms 
show the contributions from background processes. 
Signal distributions with ($m_{\lsb}$,$m_{\lntr}$)=(130,85)~GeV 
and $m_{LQ}=240$~GeV are
 shown as solid and dashed lines, respectively
}
\end{figure}

Additional selections reduce the remaining number of events with poorly 
measured $\etmiss$. We require 
${\Delta\phi}_{\text{min}}(\etmiss, {\rm jets})>0.6$~rad, 
and define an asymmetry
${\cal A} = (\etmiss-\htmiss)/(\etmiss+\htmiss)$ and require
$-0.1 < {\cal A} <0.2$~\cite{zhold}.
The $\etmiss$ and $H_T$ distributions
after imposing $b$-tagging and the requirements on 
${\Delta\phi}_{\text{min}}(\etmiss,{\rm jets})$ and ${\cal A}$
 are shown in Fig.~\ref{fig:L3VTbtag}, along with the expectations for two 
possible signals  which show the kinematic variation
for different masses. 

\begin{table*}[th]
  \caption{\label{table-jlip-sl}Predicted and observed numbers of events 
before and after $b$-tagging
and additional event selections. 
The number of 
background events after pretag selection is normalized to the number of 
data events.
Signal acceptances and the predicted number of events are given for two 
($m_{\lsb}$,$m_{\lntr}$) mass points. The acceptances for 
$m_{LQ}=240$~GeV and ($m_{\lsb}$,$m_{\lntr}$)=(240,0)~GeV are identical.
The uncertainties on the total
background and the signals include all statistical and systematic
uncertainties.
}

  \begin{ruledtabular}
  \begin{tabular}{lccccc}
       Process                         &  Pretag            &  $b$-tag            
& $-0.1<{\cal A}<0.2$                   &  $X_{jj}>0.75$          & $X_{jj}>0.9$           \\
                                       &                    &                     
& ${\Delta\phi}(\etmiss,{\text {jets}})>0.6$ &  $p^{\text{jet1}}_T 
>20$~GeV & $p^{\text{jet1}}_T >50$~GeV \\
                                       &                    &                     &                                      &  $\etmiss>40$~GeV     & $\etmiss>150$~GeV    \\
                                       &                    &                     &                                      &  $H_T>60$~GeV         & $H_T>220$~GeV        \\
    \hline
    Diboson                            &  \phantom{0}\phantom{0}2,060              
&  
\phantom{0}\phantom{0}\phantom{0}38              
&  \phantom{0}\phantom{0}\phantom{0}35                               & 
\phantom{0}31                    
& 0.3                  \\
$W(\rightarrow l{\nu})$ + light jets   &  \phantom{0}49,250             &  
\phantom{0}\phantom{0}130              
&  \phantom{0}\phantom{0}119                               & 105                   
& 
0.5                  \\
    $Wc\bar{c}, Wb\bar{b}$             &   \phantom{0}\phantom{0}7,792              
&  
\phantom{0}\phantom{0}353              
&  \phantom{0}\phantom{0}325                               & 261                   
& 
1.9                  \\
 $Z(\rightarrow ll)$  + light jets     &  \phantom{0}17,663             &  
\phantom{0}\phantom{0}\phantom{0}11             
&  \phantom{0}\phantom{0}\phantom{0}\phantom{0}9                              
& 
~~\phantom{0}8~                  
&   
0                 \\
    $Zc\bar{c},Zb\bar{b}$              &   \phantom{0}\phantom{0}4,526              
&  
\phantom{0}\phantom{0}256              
&  \phantom{0}\phantom{0}247                               & 217                   
& 
1.9                 \\
    Top                                &   \phantom{0}\phantom{0}2,019              
&  
\phantom{0}\phantom{0}348              
&  \phantom{0}\phantom{0}301                               & 190                   
& 
2.2                  \\
    MJ                                 &  \phantom{0}30,243             &  
\phantom{0}\phantom{0}444              
&  \phantom{0}\phantom{0}205                               & 157                   
&    0                  \\
\hline
Total background                       & 113,553            &  1,579 $\pm$ 230   &  1,242 $\pm$ 188                      & 971 $\pm$ 152          &  6.9 $\pm$ 1.7        \\
     \hline
\# data events                         & 113,553            &  ~1,463             
&  ~1,131                               & 901                    &       7               
\\
    \hline
Signal (acceptance, \%)               &                     &                     &                                     &                        &                       \\
($m_{\lsb}$,$m_{\lntr}$)=(240,0)~GeV  & ~145 $\pm$ 11~(38.7)  & 
~~~~43.3 $\pm$ 6.4 (11.4) &  ~~42.0 $\pm$ 6.2 (11.1)  & ~-                &  10.5 $\pm$ 1.9 (2.8) \\
($m_{\lsb}$,$m_{\lntr}$)=(130,85)~GeV & ~1,928 $\pm$ 158 ~(10.9)  & ~~~~544 $\pm$ 85 ~~(3.1)  &  ~~~529 $\pm$ 77 ~~(3.0)   & ~~~~~~~481 $\pm$ 66 (2.7) & - ~~~~               \\
   \end{tabular}
\end{ruledtabular}

\end{table*}

We then apply final selections to improve the sensitivity.
As our signals consist of two high-$p_T$ $b$-jets,
we use  $X_{jj} \equiv  ( p_{T}^{\text{jet1}} +  p_{T}^{\text{jet2}} 
)/H_T$ 
 as a discriminant against top-quark processes.
We optimize selections on $p_{T}^{\text{jet1}}$, $\etmiss$,
$H_{T}$, and $X_{jj}$ 
for different ($m_{\lsb},m_{\lntr}$) and  $m_{LQ}$ by 
choosing 
selections that yield the smallest expected limit on the cross section.  
These selections are more restrictive for  
$LQ_3$ and $\lsb$ signals with larger mass. For regions
with small $m_{\lsb}-m_{\lntr}$, the average $\etmiss$ and jet 
energies are lower, and relaxed requirements are found to be optimal. 
The results of the selections, and the predicted numbers of events 
from background processes are listed in 
Table~\ref{table-jlip-sl}, including 
two final signal selection examples. For a signal with high $\etmiss$,
the largest backgrounds are from $W/Z$ + $b\bar{b}$ production 
and top quark processes. 
There is in addition a significant contribution from multijets for bottom 
squark signal points with a 
small value of $\etmiss$. 

Systematic uncertainties include those 
on the integrated luminosity (6.1\%), 
trigger efficiency (2\%), and jet energy calibration and 
reconstruction (3\% for signal and (2--7)\% for background).
Uncertainties associated with $b$-tagging 
are (6--17)\% for signal
and (5--11)\% for background.  Uncertainties on 
theoretical cross 
sections for SM processes include 10\% on top quark production, and
6\% on the total ($W/Z$)+jets cross section with an additional 
20\%
uncertainty on heavy flavor content. The contribution from the MJ 
background
 is assigned a 25\% uncertainty which includes the impact of
possible signal events contained in the pretag sample.

We obtain limits on the pair production cross section multiplied by the 
branching fraction squared ($\sigma\times B^2$) 
using the $CL_s$ approach \cite{Junk}. In this technique, an 
ensemble of MC experiments using 
the expected numbers of signal and 
background events is compared to the number of events observed in data
to derive an exclusion limit. Signal and background 
contributions are varied within their uncertainties
taking into account correlations among their systematic uncertainties.
The $LQ_3$ and $\lsb$ ($m_{\lntr}=0$) observed and expected cross 
section limits are given in 
Table~\ref{tab::95CL_LQ3}.

\begin{table}[th]
\caption{\label{tab::95CL_LQ3} Observed and expected 95\% C.L. limits 
on the cross 
section for different leptoquark or bottom squark (assuming $m_{\lntr}=0$) 
masses.}
\begin{ruledtabular}
\begin{tabular}{lccccc}
   Mass (GeV)    &  220     & 240   & 250    & 260    & 280  \\
 \hline
   Observed (pb)    &  0.077   & 0.063 & 0.056  & 0.052  & 0.054\\
   Expected (pb)    &  0.067   & 0.056 & 0.049  & 0.046  & 0.040\\
\end{tabular}
\end{ruledtabular}
\end{table}

\begin{figure*}[th]
\begin{center}
   \includegraphics[scale=.42]{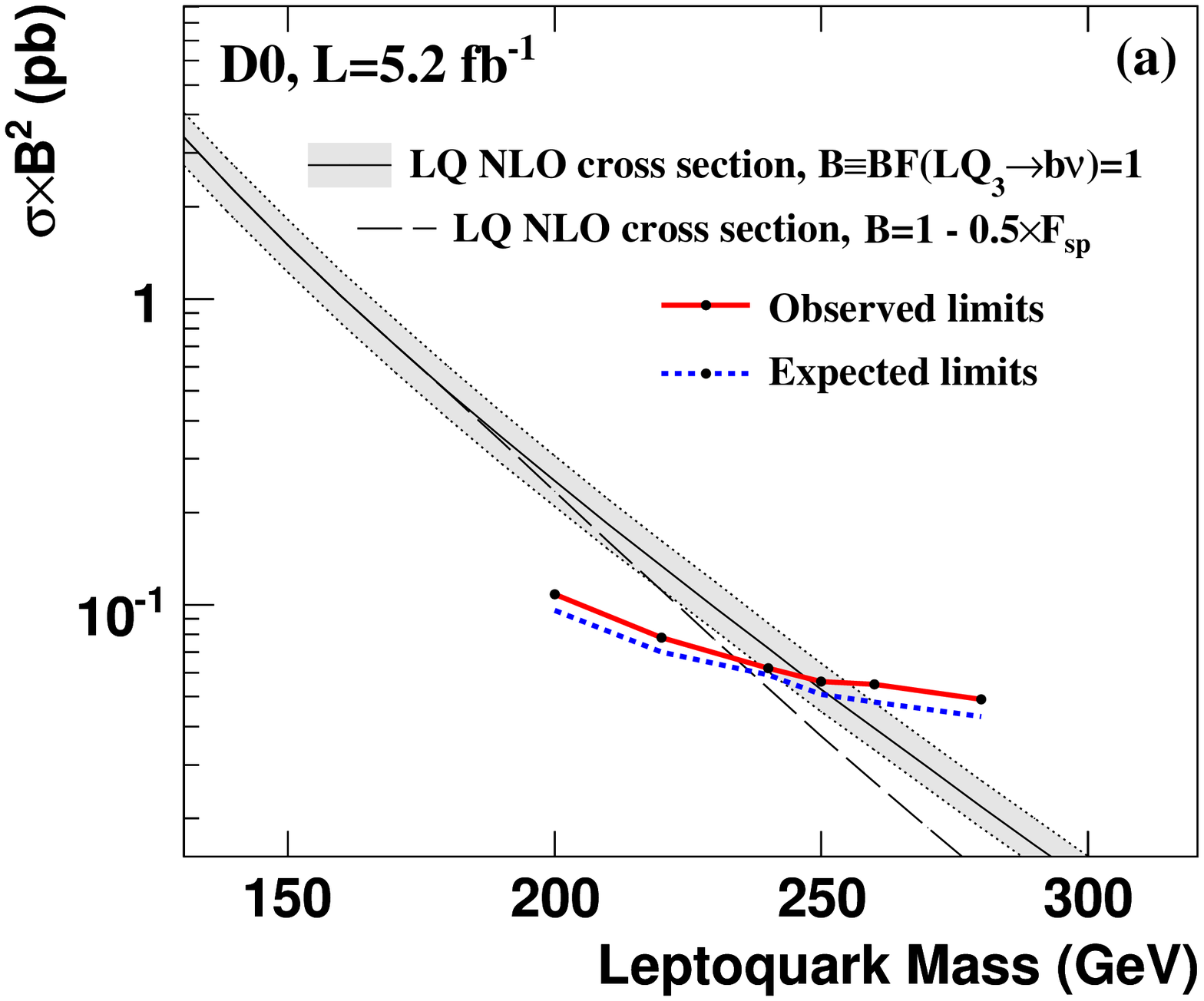}
   \includegraphics[scale=.42]{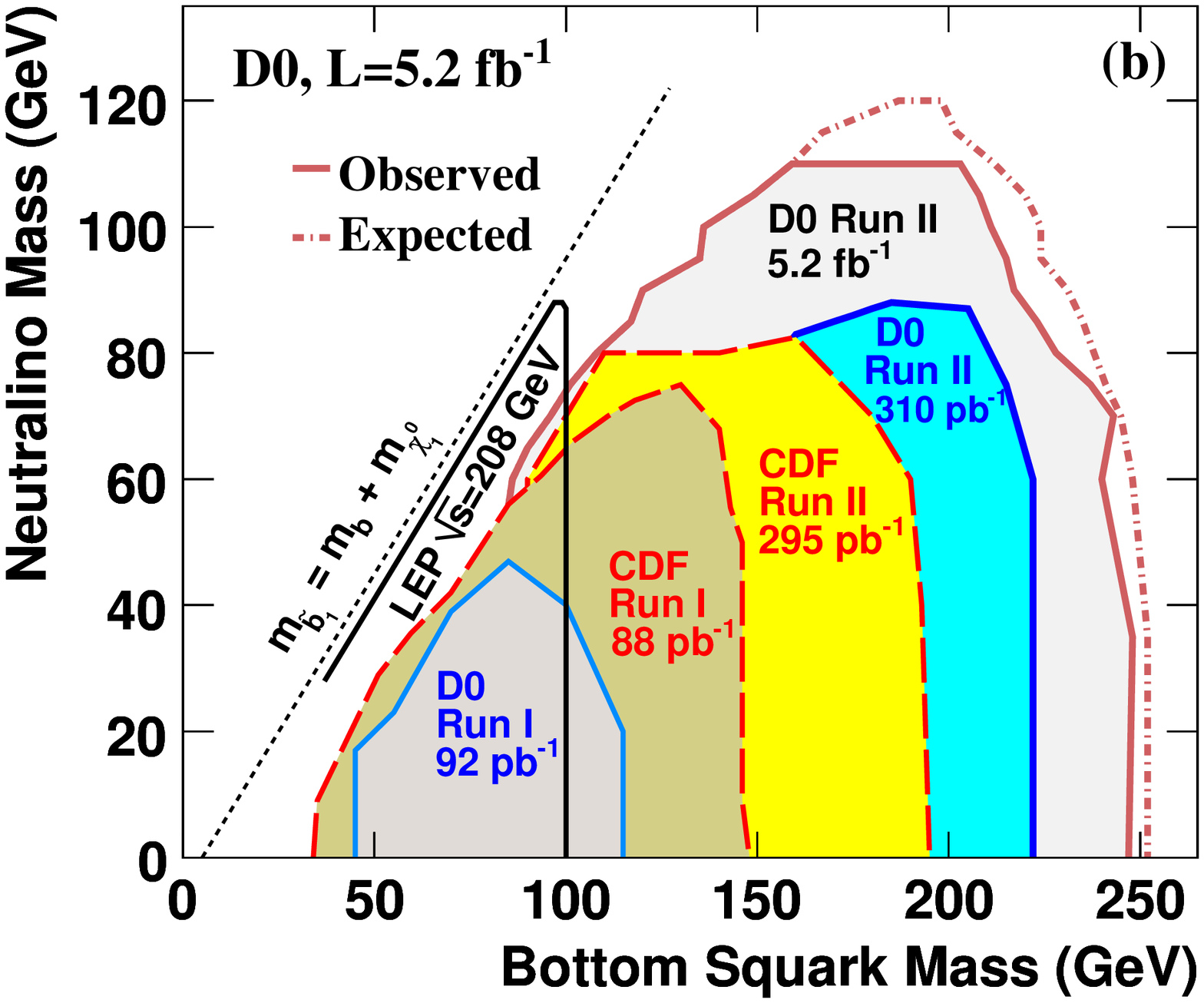}
\end{center}
   \caption{\label{fig:Lq3_Exclusion_final} (color online). (a) The 
95\%~C.L. expected (dashed line) and observed (points plus solid 
line) limits  on 
$\sigma \times B^2$  as a function of $m_{LQ}$ for the pair production 
 of third-generation leptoquarks where $B$ is the branching fraction
to $b\nu$. 
The theory band is
 shown in grey with an uncertainty range as discussed in the text.
 The long-dashed line indicates the
expected suppression of
$\sigma \times B^2$ above the $t\tau $ threshold for equal $b\nu $ and 
$t\tau $ couplings.
 (b) The 95\%~C.L. exclusion contour in the ($m_{\lsb}, m_{\lntr}$) 
plane. 
  Also shown are results from previous searches at LEP~\cite{sboldLEP} and 
the Tevatron~\cite{sboldD0,sboldCDF}.  }
 \end{figure*}

Figure~\ref{fig:Lq3_Exclusion_final}(a) shows the 95\%~C.L. upper 
limits on the cross 
section as a function of $m_{LQ}$, 
together with
the theoretical cross section $\sigma_{\text {th}}$ assuming $B=1$. The 
uncertainty on $\sigma_{\text {th}}$ is 
obtained by varying the renormalization 
and factorization scales 
by a factor of two from the nominal choice $\mu = m_{LQ}$ and
incorporating the PDF
uncertainties~\cite{Kramer:1997hh}. Limits on 
$m_{LQ}$ are obtained 
from the intersection of 
the observed cross section limit with the central $\sigma_{\text {th}}$
and yield a lower mass
limit of 247~GeV for $B=1$ for the production of 
third-generation leptoquarks.  
If the 95\% C.L.
experimental limit is compared with the 
one standard deviation lower value of $\sigma_{\text {th}}$,
we obtain a mass limit of 
$m_{LQ}=238$~GeV.
Also shown is the central value of $\sigma_{\text {th}}$
when the coupling to the $b\nu$ and $t\tau $ channels
are identical, yielding $B=1-0.5\times F_{sp}$ where $F_{sp}$ is a phase 
space suppression factor for the $\tau t$ channel~\cite{lqold}. The 
mass limit in this case is 234~GeV. 

Figure~\ref{fig:Lq3_Exclusion_final}(b) shows the  
excluded region in the plane of the
bottom squark versus neutralino mass obtained using the
central $\sigma_{\text {th}}$. For 
$m_{\lntr}=0$, the
limit is $m_{\lsb}>247$~GeV. The exclusion region extends 
to $m_{\lntr}=110$~GeV for $160<m_{\lsb}< 200$~GeV.

In conclusion, in the 5.2~fb$^{-1}$ data sample studied, 
the observed number of events with 
the topology of two $b$-jets plus missing transverse energy
 is consistent with that expected from 
known SM processes. We set limits on the
cross section multiplied by square of the branching fraction $B$ 
 to the $b\nu$
final state as a function of leptoquark mass. These results are 
interpreted as mass limits
and give a limit of 247~GeV for $B=1$ for the production of
charge-1/3 third-generation scalar leptoquarks. We also exclude the
production of bottom squarks for a range of values in the
($m_{\lsb}, m_{\lntr}$) mass plane
 such as
 $m_{\lsb}>247$~GeV for $m_{\lntr}=0$ and $m_{\lntr}>110$~GeV for
$160<m_{\lsb}< 200$~GeV. These 
limits significantly extend
previous results.

%
%

%
%
We thank the staffs at Fermilab and collaborating institutions,
and acknowledge support from the
DOE and NSF (USA);
CEA and CNRS/IN2P3 (France);
FASI, Rosatom and RFBR (Russia);
CNPq, FAPERJ, FAPESP and FUNDUNESP (Brazil);
DAE and DST (India);
Colciencias (Colombia);
CONACyT (Mexico);
KRF and KOSEF (Korea);
CONICET and UBACyT (Argentina);
FOM (The Netherlands);
STFC and the Royal Society (United Kingdom);
MSMT and GACR (Czech Republic);
CRC Program and NSERC (Canada);
BMBF and DFG (Germany);
SFI (Ireland);
The Swedish Research Council (Sweden);
and
CAS and CNSF (China).
\end{document}